\documentclass[aps,prc,twocolumn,floatfix,nofootinbib,superscriptaddress]{revtex4-1}
\usepackage{amsmath}
\usepackage{amsfonts}
\usepackage{amssymb}
\usepackage{bm}
\usepackage{graphicx}
\usepackage{multirow}
\usepackage{color}
\usepackage{mathrsfs}
\usepackage{soul}
\usepackage{bbold}
\newcommand{\be}{\begin{equation}}
\newcommand{\ee}{\end{equation}}
\newcommand{\beq}{\begin{equation}}
\newcommand{\eeq}{\end{equation}}
\newcommand{\bea}{\begin{eqnarray}}
\newcommand{\eea}{\end{eqnarray}}
\newcommand{\nn}{\nonumber}
\usepackage[colorlinks=true,linkcolor=blue,citecolor=blue,urlcolor=blue]{hyperref}

\begin{document}

\title{Charmed mesons with a symmetry-preserving contact interaction}

\author{Fernando~E. Serna}
\affiliation{Instituto de F\'{\i}sica Te\'{o}rica, Universidade Estadual
Paulista, Rua Dr. Bento Teobaldo Ferraz, 271 - Bloco II, 01140-070 S\~{a}o Paulo, SP, Brazil}

\author{Bruno El-Bennich}
\affiliation{Laboratorio de F\'{i}sica Te\'{o}rica e Computacional, Universidade Cruzeiro do Sul, 
Rua Galv\~{a}o Bueno 868, 01506-000 S\~{a}o Paulo SP, Brazil}

\author{\mbox{Gast\~{a}o~Krein}}
\affiliation{Instituto de F\'{\i}sica Te\'{o}rica, Universidade Estadual
Paulista, Rua Dr. Bento Teobaldo Ferraz, 271 - Bloco II, 01140-070 S\~{a}o Paulo, SP, Brazil}

\begin{abstract}
A symmetry-preserving treatment of a vector-vector contact interaction is used to study charmed 
heavy-light mesons. The contact interaction is a representation of nonperturbative kernels used 
in Dyson-Schwinger and Bethe-Salpeter equations of QCD. The Dyson-Schwinger equation is solved
for the $u,\,d,\,s$ and $c$ quark propagators and the bound-state Bethe-Salpeter amplitudes respecting 
spacetime-translation invariance and the Ward-Green-Takahashi identities associated with global 
symmetries of QCD are obtained to calculate masses and electroweak decay constants of the pseudoscalar 
$\pi,\,K$, $D$ and $D_s$ and vector $\rho$, $K^*$, $D^*$, and $D^*_s$ mesons. The predictions of the 
model are in good agreement with available experimental and lattice QCD data. 
\end{abstract}


%

\maketitle
\section{Introduction}
\label{sec:intro}

Heavy-light $Q\bar q$ (and $\bar Q q$) mesons, such as the $B$ and the~$D$, are interesting bound 
states of quantum chromodynamics (QCD). They are interesting because they are composed of quarks 
belonging to two limiting mass sectors of QCD with associated emergent approximate symmetries: 
the sector of light quarks $q = (u,d,s)$, with masses $m_q \ll \Lambda_{\rm QCD}$, and the sector of 
heavy quarks $Q=(c,b,t)$, with masses $m_Q \gg \Lambda_{\rm QCD}$, where $\Lambda_{\rm QCD}$ is the 
energy scale at which the theory becomes strongly coupled, thereby implying that the characteristic 
size of a typical hadron is $\Lambda^{-1}_{\rm QCD}$. In the $m_q \rightarrow~0$ limit, QCD 
acquires an $SU(3)_L \times SU(3)_R$ chiral symmetry that is dynamically broken by the strong QCD 
interactions to an $SU(3)_V$ flavor symmetry; the eight pseudoscalar mesons $\pi^0$, $\pi^{\pm}$, 
$K^0$, ${\bar K}^0$, $K^{\pm}$, and $\eta$ are identified with the (pseudo-)Goldstone bosons associated 
with the dynamical breaking of the symmetry. In the $m_Q \rightarrow \infty$ limit, the interactions of 
a heavy quark, regardless of its flavor, within a heavy-light meson become independent of its spin, a feature
that gives rise to a spin-flavor heavy quark $U(2N_h)$ symmetry{\textemdash}$\,2N_h = 2\,({\rm spin})\times N_h\,
(\rm{heavy\;flavors})$. Moreover, since the average velocity $v$ of the heavy quark in a $Q\bar q$ bound 
state is changed very little by the interactions, as $\Delta v = \Delta p/m_Q \sim \Lambda_{\rm QCD}/m_Q 
\ll 1$, the light quark dynamics occurs in the background of a strong color field of an essentially static 
spectator. Therefore, heavy-light mesons offer a unique opportunity to learn about features of dynamical 
chiral symmetry breaking (D$\chi$SB) in a spin- and flavor-independent environment provided by the 
heavy quark. 

In fact, the approximate chiral and heavy quark spin-flavor symmetries can be combined to construct 
powerful effective field theories to make predictions for a wealth of processes involving heavy-light mesons, 
like electroweak decay rates and their low-energy interactions with other 
hadrons~\cite{{Manohar:2000dt},{Petrov:2016azi}}. However, as with many other effective field theories, 
coupling constants in the Lagrangian of chiral heavy-quark effective field theories, and also 
form factors associated with decay matrix elements, are of nonperturbative origin and need be fixed from 
data or from other theoretical source, as simulations of QCD on a space-time lattice or calculations 
using nonperturbative methods in the continuum. The present work is related to the latter;
in particular, to an approach based on the Dyson-Schwinger (DS) and Bethe-Salpeter (BS) equations of 
QCD~\cite{{Roberts:1994dr},{Alkofer:2000wg}}.

The DS and BS equations consist of an infinite set of coupled integral equations; once a truncation scheme 
is specified, they define a tractable and predictive problem. Systematic, symmetry-preserving, 
nonpertubatively renormalizable truncation schemes, continuously developed since the 1990, reached a high 
degree of sophistication and have proven very successful in describing and correlating a great variety of 
phenomena in the light-quark sector of QCD~\cite{Cloet:2013jya,{Eichmann:2016yit}}. Symmetry-preserving 
schemes make use, in an essential way, of the Ward-Green-Takahashi (WGT) identities reflecting global 
symmetries and their explicit breaking; they impose stringent relationships between the interaction kernels 
entering DS e\-qua\-tions for quark and gluon propagators and quark-gluon vertices and those entering BS 
equations for bound states~\cite{Binosi:2016rxz}. 

Notwithstanding the advances and successes, challenges 
still remain in describing simultaneously the masses and decay constants of light- and heavy-flavored 
mesons within a single interaction-truncation scheme~\cite{{Maris:2005tt},{Nguyen:2010yh},
{Souchlas:2010boa},{Souchlas:2010zz},{Bashir:2012fs},{Gomez-Rocha:2014vsa},{Rojas:2014aka},
{Gomez-Rocha:2015qga},{Gomez-Rocha:2016cji},{Hilger:2017jti}}; in particular, the disagreements with data
for the electroweak decay constants are substantial~\cite{{Nguyen:2010yh},{Hilger:2017jti}}.  Although 
one can expect that the challenges will be overcome in a foreseeable future, there is pressing need for 
different pieces of information on the structure and 
interactions of such mesons for guiding new experiments at existing and forthcoming facilities, aiming at 
e.g. production of exotic hadrons like the X,Y,Z hadrons in heavy-ion collisions and creation of exotic 
nuclear bound states with charmed hadrons~\cite{{Briceno:2015rlt},{Krein:2016fqh}}. Predictions for masses, 
strong couplings, decay rates, and interaction cross sections are needed as functions of external parameters 
like temperature, baryon density and magnetic field, delivered in a form that can be used efficiently in transport 
and hydrodynamic simulation codes of such complex experiments. Given these circumstances and demands, 
in the present paper we explore the effectiveness in describing properties of $D$ mesons of a simpler 
alternative based on a four-fermion contact interaction (CI) model embedded in a symmetry-preserving 
scheme~\cite{Serna:2016kdb}. 

Fermionic contact interactions find widespread applications in hadron physics as evidenced by the 
popular use of models inspired by the Nambu-Jona-Lasinio (NJL)~\cite{Nambu:1961tp}. A~great deal of 
qualitative insight on the phenomenon of hadron mass generation via D$\chi$SB and 
the role of the $\pi^0$, $\pi^{\pm}$, $K^0$, ${\bar K}^0$, $K^{\pm}$, and $\eta$ mesons as the 
associated (pseudo-)Goldstone bosons has been gleaned from such models{\textemdash}for reviews, see 
Refs.~\cite{{Vogl:1991qt},{Klevansky:1992qe},{Hatsuda:1994pi},{Bijnens:1995ww}}. On the other hand, 
the lack of confinement and non-renormalizabilty are the major weaknesses of these CI models. The 
non-renormalizability, notably, carries along the danger of introducing gross violations of global symmetries 
due to the regularization procedure; ambiguities arising from momentum shifts in divergent integrals
and severe dependence of results on choices of momentum sharing between a heavy and a light quark in a
bound state are the main causes of the problems. 

A new perspective, however, has recently emerged with the 
implementation by Guti\'errez-Guerrero, Bashir, Clo\"et and Roberts~\cite{{GutierrezGuerrero:2010md}} 
(GBCR henceforth) of a confining, symmetry-preserving treatment of a vector-vector CI as a simplified 
{\em ansatz\/} for the gluon's two-point Schwinger function commonly employed in the kernel of the quark's 
DS equation~\cite{Krein:1990sf,Roberts:1994dr,Fischer:2003rp,Bashir:2012fs}. By introducing a mechanism 
that ensures the absence of quark production thresholds~\cite{Ebert:1996vx}, a feature of a  confining  theory, 
and embedding the interaction in a global-symmetry-preserving, rainbow-ladder (RL) truncation framework of the 
DS and BS equations~\cite{{Munczek:1994zz},{Bender:1996bb}}, the GBCR scheme has enhanced the CI's capacity 
to describe in a unified manner a diverse array of phenomena that include the light-quark meson and baryon 
spectra as well as their electroweak, elastic and transition form factors~\cite{{Roberts:2011wy},
{Roberts:2010rn},{Chen:2012qr},{Wilson:2011aa},{Chen:2012txa},{Roberts:2011cf},{Wang:2013wk},
{Segovia:2013rca},{Segovia:2013uga}} in appropriate kinematic regimes. Very recently, the scheme has been 
extended to calculate ground-state masses and weak decay constants 
of heavy charmonia~\cite{Bedolla:2015mpa}, and also elastic and transition form factors of 
$\eta_c(1S)$~\cite{Bedolla:2016yxq}.

In this paper, we examine the GBCR approach within the perspective of a subtraction scheme that allows 
one to isolate symmetry-violating contributions in 
BS amplitudes for arbitrary momentum routing in occurring divergent integrals. The scheme has 
been employed in  NJL model calculations in vacuum~\cite{Battistel:2008fd,Battistel:2013cja} 
and at finite temperature and baryon density~\cite{{Battistel:2003gn},{Farias:2006cs},Farias:2005cr,Farias:2007zz}, and very recently to investigate the critical behavior of quark matter in presence of 
a chiral imbalance~\cite{Farias:2016let}. The scheme is inspired in the method introduced in 
Ref.~\cite{Batt-thesis}, the aim of which is the treatment of divergent Feynman integrals without 
specification of an explicit regulator; it shares similarities~\cite{Sampaio:2002ii} with the Bogoliubov, 
Parasiuk, Hepp, Zimmermann (BPHZ) renormalization subtraction scheme~\cite{Collins:1984xc}, which uses 
systematic subtractions of momentum space integrals to isolate divergences. 

We here take advantage of the strengths of the subtraction scheme and apply it to heavy-light mesons 
within the CI~approach. Heavy-light mesons were studied previously in the NJL model using a traditional 
cutoff regularization~\cite{blaschke}. However, a wrong pattern in the ordering of the pseudoscalar 
decay constants, $f_D < f_\pi$ was found{\textemdash}the experimental pattern is $f_\pi < f_K < f_D$.
More recently, a similar incorrect ordering in the decay constants was encountered within the CI~framework 
for heavy charmonia in Ref.~\cite{Bedolla:2015mpa}.  In addition, as already mentioned, even implementing 
a more realistic interaction in the RL truncation of the DS and BS equations, the weak decay constants of 
heavy-light mesons compare unfavorably~\cite{Nguyen:2010yh} with predictions of lattice-QCD and the $D$ to 
$D_s$ mass difference is vanishingly small~\cite{Rojas:2014aka}. Possible causes for the failure of this 
truncation in heavy-light systems have been put forward in Ref.~\cite{El-Bennich:2016qmb}. While not 
being a substitute for a full-fledged QCD-based DS-BS framework currently under intense development, 
it is nonetheless legitimate to expect that the capacity of a CI scheme in providing useful insight 
on heavy-light mesons is enhanced when it respects fundamental spacetime and internal symmetries of~QCD.
 
This paper is organized as follows: in Sec.~\ref{sec:eqsWGT} we summarize the essentials of DSE, BSE and 
Ward-Green-Takahashi identities (WGTI) and introduce the CI scheme employed herein. In Sec.~\ref{secIII}, 
we introduce the subtraction scheme to deal with divergent integrals that occur in the Bethe-Salpeter kernels 
due to the simplification of the CI and discuss the consequences of the axialvector WGTI for these divergences. 
In Sec.~\ref{secIV} we present our numerical results for masses and weak decay 
constant of the mesons of interest using the subtraction scheme introduced in Sec.~\ref{secIII} 
We discuss our results and compare them with those obtained in Ref.~\cite{Bedolla:2015mpa}. 
Finally, in Sec.~\ref{secV} we conclude with some final remarks about this work.

\section{DS and BS Equations and WGT identities}
\label{sec:eqsWGT}

We begin with a brief review of the basic elements of the GBCR contact-interaction 
scheme~\cite{GutierrezGuerrero:2010md}. We consider the inhomogeneous 
BS equation for a quark and antiquark state of total momentum $P$ (here and in the following we 
omit renormalization constants):  
\begin{eqnarray}
\hspace{-0.35cm}
\left[\Gamma_{\cal M}(k;P)\right]_{AB} &=& {\cal M}_{AB} 
+ \int_q \left[K(k,q;P)\right]_{AC,DB} \nonumber \\
&&\times \, \left[S(q_+) \Gamma_{\cal M}(q;P) S(q_-)\right]_{CD},
\label{eq:inhBSE}
\end{eqnarray}
where $\int_q\equiv\int d^4 q/(2\pi)^4$,  ${\cal M}$ represents the Dirac spinor structure of the state, 
$K(q,k;P)$ is the fully amputated quark-antiquark scattering kernel; $A,B,\cdots $ denote 
collectively color, flavor, and spinor indices; $q_\pm = q  \pm \eta_\pm P$, 
with $\eta_+ + \eta_- = 1$ and $q$ is the relative momentum. $S(k)$ is the dressed-quark 
propagator given by a DSE; for a given flavor~$f$ the general form of this DSE is (in Euclidean metric)
\begin{eqnarray}
S^{-1}_f(k) &=& i\gamma\cdot k + m_f  \nonumber \\
&& + \,  \int_q \, g^2 D_{\mu\nu}(k-q) \,\frac{\lambda^a}{2}\gamma_\mu 
S_f(q) \, \Gamma^{a f}_\nu(q,k)\, , \hspace{4mm}
\label{eq:DSEqp}
\end{eqnarray}
where $m_f$ is the current-quark mass. In here, we are  interested in the flavor-nonsinglet
axial-vector $\Gamma^{lh}_{5\mu}(k;P)$ and pseudoscalar $\Gamma^{lh}_{\rm PS}(k;P)$
amplitudes for a quark-antiquark pair of a light ($l$) and a heavy ($h$) quark, with  
${\cal M}_{5\mu} = \gamma_5 \gamma_\mu$ and ${\cal M}_{PS} = \gamma_5$
respectively. Spacetime-translation invariance requires that no observable can depend on the 
choice of the momentum routing in quark propagators in Eq.~(\ref{eq:inhBSE}); that is, 
physical results must be independent of~$\eta_\pm$. 

Associated with $\Gamma^{lh}_{5\mu}$ is the WGT identity.
\begin{eqnarray}
P_\mu \Gamma_{5\mu}^{lh}(k;P)  & = &  S_l^{-1}(k_+) i \gamma_5 +  i \gamma_5 S_h^{-1}(k_-) 
\nonumber \\
&& -   \, i\,(m_l + m_h) \,\Gamma^{lh}_{\rm PS}(k;P)\,,
\label{eq:avWGT}
\end{eqnarray}
where $\Gamma^{lh}_{PS}(k;P)$ is the pseudoscalar vertex; both $\Gamma_{5\mu}^{lh}(k;P)$ and
$\Gamma^{lh}_{\rm PS}(k;P)$ obey inhomogeneous BS equation (\ref{eq:inhBSE}). Pseudoscalar 
meson bound states are obtained from the solution of the homogeneous equation for 
$\Gamma^{lh}_{\rm PS}(k;P)$
\begin{eqnarray}
\left[\Gamma^{lh}_{\rm PS}(k;P)\right]_{AB} &=& \int_q\left[K^{lh}(k,q;P)\right]_{AC,DB} 
\nonumber \\ 
&& \times \, \left[S_l(q_+)\Gamma^{lh}_{\rm PS}(q;P)S_h(q_-)\right]_{CD} ,
\label{eq:BSEps}
\end{eqnarray}
where here $A,B, \cdots$ denote color and spinor indices only. The general form of the of 
$\Gamma^{lh}_{\rm PS}(k;P)$ is
\begin{eqnarray}
\Gamma^{lh}_{\rm PS}(k,P) & = & \gamma_5 \bigl[ i E^{lh}_{\rm PS}  + \gamma\cdot P \, F^{lh}_{\rm PS} 
+ \gamma\cdot k \, G^{lh}_{\rm PS} 
\nonumber \\
&& + \,  \sigma_{\mu\nu} k_\mu P_\nu \, H^{lh}_{\rm PS} \bigr ],
\label{GammaPS}
\end{eqnarray}
where $E^{lh}_{\rm PS}, F^{lh}_{\rm PS}, \cdots$ are functions of $k$, $P$ and $k\cdot P$. 
The meson mass, $m_{\rm PS}$, is the eigenvalue for the value $P^2 = - m^2_{\rm PS}$ that 
solves Eq.~(\ref{eq:BSEps}). 

The CI scheme introduced in~Ref.~\cite{GutierrezGuerrero:2010md} amounts to the following 
replacement in Eq.~(\ref{eq:DSEqp}) 
\begin{eqnarray}
g D_{\mu\nu}(k-q) \, g \Gamma^{a f}_\nu(q,k) &\rightarrow& 
\left( \frac{4\pi\alpha_{\rm IR}}{m^2_{g}}\right)^f 
\frac{\lambda^a}{2}\gamma_\mu 
\nn \\
& \equiv & \left(\frac{1}{m^f_{\rm G}}\right)^2  \, \frac{\lambda^a}{2}\gamma_\mu ,
\label{eq:contact}
\end{eqnarray}
where $m_{g}$ is a gluon mass-scale and $\alpha_{\rm IR}$ is a coupling strength parameter. 
Note that a flavor dependence in the interaction strength is due to the flavor dependence of 
the full quark-gluon vertex $\Gamma^{a f}_\mu$. The flavor dependence is important to accommodate 
the fact that heavy-flavor quarks probe shorter distances than light-flavor quarks at the 
corresponding quark-gluon vertices, thereby implying a smaller coupling strength for heavy-flavor 
quarks~\cite{El-Bennich:2016qmb}. This fact was used previously in the NJL model to mimic the short-distance 
and weak-coupling physics in quark matter at high temperatures and baryon densities~\cite{{Casalbuoni:2003cs},
{Farias:2006cs}}. Very recently, Refs.~\cite{{Bedolla:2015mpa},{Bedolla:2016yxq}} have shown that good 
agreement with experiment for heavy-charmonia masses can be obtained by using a weaker coupling strength 
and  a simultaneous increase in the ultraviolet cutoff of the CI of Ref.~\cite{GutierrezGuerrero:2010md}. 
We anticipate that the same turns out to be relevant for heavy-light mesons and write for the BS kernel
\begin{equation}
\hspace{-0.28cm}\left[K^{lh}(k,q;P)\right]_{\!AC,DB} =  - 
\frac{1}{m^l_{\rm G} m^h_{\rm G}}\left(\frac{\lambda^a}{2}\gamma_\mu\right)_{\!\!\!AC} 
\!\!\left(\frac{\lambda^a}{2}\gamma_\mu\right)_{\!\!\!DB}\!\!\!.
\label{eq:Kcontact}
\end{equation}

A feature of the momentum independence of the CI is that the corresponding DS equation in
Eq.~\eqref{eq:DSEqp} is non-renormalizable. In addition, the BS equation in the CI acquires ultraviolet 
divergencies and is also non-renormalizable. This implies that mass-scale parameters introduced 
with the regularization of divergent integrals cannot be removed from the calculations and need 
to be fixed phenomenologically. Another feature, as previously  mentioned, concerns the regularization 
of divergent integrals: they carry along the danger of symmetry violation, in particular the WGTI in 
Eq.~(\ref{eq:avWGT}) is not satisfied even when Poincar\'e-invariant regularization schemes are employed. 

Let us consider the DS and the homogeneous pseudoscalar BS equations, Eqs.~(\ref{eq:DSEqp})
and~(\ref{eq:BSEps}) respectively, with the CI approximation. In this case the solution of the gap equation 
becomes, $S^{-1}_f(k)=  i\gamma\cdot k + M_f$, with the momentum independent quark-mass function,
\begin{eqnarray}
\label{gapequ}
M_f = m_f + \frac{16}{3}\left(\frac{1}{m^f_{\rm G}}\right)^2 \, 
\int_q\, \frac{M_f}{q^2+M^2_f} . 
\label{gapMf}
\end{eqnarray}
It is remarkable that the non-running quark mass, $M_f$, is related to the commonplace constituent quark 
masses of quark and light-front models~\cite{ElBennich:2008xy,daSilva:2012gf,ElBennich:2012ij,deMelo:2014gea,
Yabusaki:2015dca}, yet it is dynamically generated. Such a momentum independent mass function is appropriate 
in the calculation of static observables, such as the meson mass spectrum, weak decay constants and charge radii, 
but leads to {\em hard\/} elastic and transition form factors that strongly depart from experimental data for 
$q^2 > \Lambda^2_\mathrm{QCD}$~\cite{Roberts:2011wy,GutierrezGuerrero:2010md,Roberts:2010rn,Chen:2012qr,
Wilson:2011aa,Chen:2012txa,Roberts:2011cf,Wang:2013wk,Segovia:2013rca,Segovia:2013uga,Bedolla:2016yxq,
ElBennich:2008qa,ElBennich:2009vx}.

In this same CI framework, the pseudoscalar BS amplitude, for example, is independent of the relative 
quark-antiquark momentum.  As a consequence, $G^{lh}_{\rm PS} = H^{lh}_{\rm PS} = 0$, and 
Eq.~\eqref{GammaPS} reduces~to,
\begin{eqnarray}
\label{contactBSA}
\Gamma^{lh}_{\rm PS}(P)=\gamma_5\left [ \, iE^{lh}_{\rm PS}(P)+\frac{1}{2M_{lh}}\, 
\gamma \cdot P\,F^{lh}_{\rm PS}(P) \right ],
\end{eqnarray}
where $M_{lh}=M_lM_h/(M_l+M_h)$. Therefore, the  BS equation can be written in the matrix form,
\begin{eqnarray}
\label{MHSE_mps}
\hspace*{-0.6cm}
\left[
\begin{array}{c}
E^{lh}_{\rm PS}(P)\\[0.2true cm]
F^{lh}_{\rm PS}(P)
\end{array}
\right]
&=& \frac{1}{3m^{l}_{\rm G}m^{h}_{\rm G}}
\left[
\begin{array}{cc}
{\cal K}^{EE}_{\rm PS} & {\cal K}^{EF}_{\rm PS} \\[0.2true cm]
{\cal K}^{FE}_{\rm PS} & {\cal K}^{FF}_{\rm PS}
\end{array}
\right]
\left[\begin{array}{c}
E^{lh}_{\rm PS}(P)\\[0.2true cm]
F^{lh}_{\rm PS}(P)
\end{array}
\right],
\end{eqnarray}
where the kernel's matrix elements are given by,
\begin{eqnarray}
\hspace{-0.5cm}{\cal K}^{EE}_{\rm PS} &=& - \int_q
{\rm Tr} \left [ \gamma_5  \gamma_\mu S_l(q_+)  \gamma_5 S_h(q_-) \gamma_\mu \right ],
\label{EE} \\ 
\hspace{-0.5cm}{\cal K}^{EF}_{\rm PS} &=& \frac{i}{2M_{lh}} \int_q
{\rm Tr} \left [ \gamma_5 \gamma_\mu \, S_l(q_+) \gamma_5  {\gamma\cdot P} 
S_h(q_-)  \gamma_\mu \right ], 
\label{EF} \\ 
\hspace{-0.5cm}{\cal K}^{FE}_{\rm PS} &=& \frac{2iM_{lh}}{P^2}\int_q
{\rm Tr} \left [ \gamma_5  {\gamma\cdot P}  \gamma_\mu S_l(q_+) \gamma_5 
S_h(q_-) \, \gamma_\mu \right ] ,
\label{FE} \\ 
\hspace{-0.5cm}{\cal K}^{FF}_{\rm PS} &=& \frac{1}{P^2}\int_q
{\rm Tr}\left [ \gamma_5 {\gamma\cdot P} \gamma_\mu  S_l(q_+)\gamma_5 
 {\gamma\cdot P}  S_h(q_-) \,\gamma_\mu \right ].  
\label{FF}
\end{eqnarray}
In Eqs.~\eqref{EE}--\eqref{FF} the traces are over Dirac indices. All integrals in Eq.~(\ref{gapMf}) and 
Eqs.~(\ref{EE})--(\ref{FF}) are ultraviolet divergent; the divergences are quadratic  and logarithmic. The vast 
majority of applications within NJL models ignore the  pseudo vector component $F^{lh}_{\rm PS}(P)$; 
in doing so leads to the random-phase-approximation (RPA) of the BS equation~\cite{{Vogl:1991qt},
{Klevansky:1992qe},{Hatsuda:1994pi},{Bijnens:1995ww}}. 

For vector mesons, the corresponding BS equation in the RL CI model is given by
\bea
\Gamma^{lh}_{\rm V}(P) = \gamma^{\bot}_\mu\,E^{lh}_{\rm V}(P)~,
\eea
where 
\bea
\gamma^{\bot}_\mu = \gamma_\mu - \frac{\gamma\cdotp P}{P^2} P_\mu.
\eea
With only one Lorentz covariant the BS equation for the vector meson simplifies to
\bea
1=\frac{1}{m^l_G\,m^h_G}K^{EE}_{\rm V}(P)~,
\eea
with 
\beq
K^{EE}_{\rm V}(P) = -\frac{1}{3} \int_q 
{\rm Tr}[\gamma^{\bot}_\nu \, \gamma_\mu \, S_l(q_+) \, \gamma^{\bot}_\nu \, S_h(q_-) \, \gamma_\mu].
\label{KEE-V}
\eeq
%

\section{Symmetry-preserving subtraction scheme}
\label{secIII}

The issue of symmetry violation can be exposed examining the momenta running in the quark 
propagators in the BS amplitudes in Eqs.~(\ref{EE})--(\ref{FF}); they are $q_+ = q +\eta_+ P$ 
and $q_- = q - \eta_- P$, where $\eta_\pm$ are arbitrary partition variables satisfying 
$\eta_+ + \eta_- = 1$. However, to maintain translational invariance, the results of the integrals 
can depend on the relative momentum $q_+ - q_-$ only, or,  equivalently, they must not 
depend on $\eta_\pm$ individually but solely on the combination $\eta_+ + \eta_- = 1$. This dependence on 
the relative momentum is also crucial for preserving the WGTI in Eq.~(\ref{eq:avWGT}), as we discuss 
shortly ahead. Moreover, for very different values of $M_h$ and $M_l$, the independence
of the results on $\eta_+$ and $\eta_-$ is a serious issue in RL finite-range 
models~\cite{Gomez-Rocha:2016cji} which becomes exacerbated in nonrenormalizable CI models. 

Within NJL models, the customary way of handling integrals, such as in  Eqs.~(\ref{EE})--(\ref{FF}), 
is as follows~\cite{{Vogl:1991qt},{Klevansky:1992qe},{Hatsuda:1994pi},{Bijnens:1995ww}}: after evaluating the 
traces, a choice for $\eta_+$ and $\eta_-$ is made and Feynman parameters are used to combine in a 
single term the product $(q^2_+ + M^2_l)(q^2_- + M^2_h)$ in the denominator. Thereafter a {\em momentum shift} 
is applied to eliminate the angle defined by the scalar product, $q{\cdot}P$, in the denominator and finally the 
integral over $q$ is performed. In shifting the momentum, changes in the integration limits are 
ignored. Invariably, results depend upon the choices made for $\eta_\pm$;  in particular, the value of 
the pion decay constant $f_\pi$, sensitive to the normalization of the pion BS equation, depends on the 
choices of the partition parameters.  In some instances regularization-independent results can be obtained 
after using the gap equation to eliminate the quadratic divergences, like in the derivation of a 
Goldberg-Treiman relation at the quark level and the Gell-Mann--Oakes-Renner relationship{\textemdash}see 
e.g. the discussions around Eq.~(4.27) in Ref.~\cite{Klevansky:1992qe}. 

The subtraction scheme is based on the repeated  use of the identity,
\begin{eqnarray}
\frac{1}{q^2_\pm + M^2_{l,h}} &=& \left(\frac{1}{q^2_\pm + M^2_{l,h}} - \frac{1}{q^2 + M^2}\right) 
+ \frac{1}{q^2 + M^2} \nonumber \\ 
&=&  \frac{1}{q^2 + M^2} - \frac{(q^2_\pm - q^2 + M^2_{l,h} -M^2 )}
{\left(q^2 + M^2\right)(q^2_\pm + M^2_{l,h})} \, ,
\label{subtr-1}
\end{eqnarray}  
where $M$ is an arbitrary subtraction mass-scale parameter. This mass scale plays a similar role to
the $\mu$ scale in dimensional regularization and can be used to tune parameters of the model
when applying it to explore physics at different scales. We do not trail this interesting possibility 
in this paper and simply keep $M$ arbitrary; we comment on this further ahead in this section.

Assuming a Poincar\'e-invariant regularization 
for the integrals in Eqs.~\eqref{EE}-\eqref{FF}, subtractions are performed in each of the  propagators, 
$S_l(q_+)$ and $S(q_-)$, the number of which is dictated by the requirement that a finite integral is obtained. 
Note that while the original denominator behaves as  $1/q^2$ in the limit $q \rightarrow \infty$, the last term in 
Eq.~(\ref{subtr-1}) tends to $1/q^4$ for $q \rightarrow \infty$. We illustrate the procedure in detail for  
the ${\cal K}^{EE}_{\rm PS}(P)$ kernel. Evaluation of the trace in Eq.~(\ref{EE}) leads to, 

\begin{widetext}
\begin{eqnarray}
\label{EE-tr}
{\cal K}^{EE}_{\rm PS} &=&  16 \int^\Lambda_q \frac{q_+ \cdot q_- +M_l M_h}{(q^2_+ + M^2_l)(q^2_- + M^2_h)}
=  8\int^\Lambda_q\Biggl\{\frac{1}{q^2_+ + M^2_l} +  \frac{1}{q^2_- + M^2_h} 
-  \frac{P^2 + \left(\Delta M_{hl}\right)^2}
{(q^2_+ + M^2_l)(q^2_+ + M^2_h)}\Biggr\}, 
\end{eqnarray}
where $\Delta M_{lh} = M_l - M_h$, and $\Lambda$ denotes the ultraviolet mass scale associated with 
the regularization. By using trice the identity of Eq.~(\ref{subtr-1}), the first two terms can be rewritten 
as
\beq 
\int^\Lambda_q \frac{1}{q^2_\pm + M^2_{l,h}} =
I_{\rm quad}(M^2_{l,h})+\eta^2_\pm \, P_\mu P_\nu A_{\mu\nu}(M^2) ,
\label{Iqua-pm}
\eeq
where $A_{\mu\nu}(M^2)$ is the integral defined by
\begin{equation}
A_{\mu\nu}(M^2) = \int^\Lambda_q  
\frac{ 4q_\mu q_\nu - (q^2+M^2) \delta_{\mu\nu} }{ (q^2+M^2)^3 } ,
\label{A_munu}
\end{equation}
which is the Euclidean space counterpart of the integral $\Delta_{\mu\nu}(M^2)$ 
in Minkowski space of Refs.~\cite{Batt-thesis,Battistel:2008fd,Battistel:2013cja}. Using Eq.~(\ref{Iqua-pm}) 
and subtracting each of the denominators in the third term in Eq.~(\ref{EE-tr}), one can write 
${\cal K}^{EE}_{\rm PS}$ can be written 
as a sum of three kinds of terms: (1) a finite integral independent of $\eta_\pm$; (2) quadratic and 
logarithmically divergent integrals that are also independent of $\eta_\pm$; and (3) a symmetry violating 
term proportional to  $\eta^2_+$ and $\eta^2_-$, namely,
\begin{eqnarray}
{\cal K}^{EE}_{\rm PS} &=&  8 \Bigl\{ 
- \left[ P^2 + \left(\Delta M_{hl}\right)^2 \right] \left[ I_{\rm log}(M^2) - Z_0(M^2_l,M^2_h,P^2;M^2)\right] 
+ I_{\rm quad}(M^2_l)  + I_{\rm quad}(M^2_h)
\nn \\
&& + \, 
\left(\eta^2_+ +\eta^2_- \right)A_{\mu\nu}(M^2)
P_\mu P_\nu  \Bigr\} ,
\label{EE-fin}
\end{eqnarray}
where $Z_0(M^2_l,M^2_h,P^2;M^2)$ is the finite integral
\beq
Z_0(M^2_l,M^2_h,P^2;M^2) = \int^1_0dz\,\int^\Lambda_q \left[
\frac{1}{(q^2+M^2)^2}
- \frac{1}{(q^2 + H(z))^2}\right] 
= \frac{1}{(4\pi)^2} \int^1_0dz\, \ln \left[\frac{H(z)}{M^2}\right],
\label{Z0-def}
\eeq
\end{widetext}
where $H(z)$ is the function,
\begin{equation}
H(z) = z(1-z)P^2 - (M^2_l-M^2_h)z +  M^2_l.
\end{equation}
Here $I_{\rm log}(M^2)$ and $I_{\rm quad}(M^2)$ and are the logarithmically and 
quadratically divergent integrals,
\begin{eqnarray}
I_{\rm log}(M^2) &=& \int^\Lambda_q \, \frac{1}{(q^2+M^2)^2},
\label{Ilog}
\\[0.2true cm]
I_{\rm quad}(M^2) &=& \int^\Lambda_q \,  \frac{1}{q^2+M^2}. 
\label{Iquad}
\end{eqnarray}
In the derivation of Eq.~(\ref{EE-fin}), we made use of the identity
\bea
&& I_{\rm quad}(M^2_{l,h}) = I_{\rm quad}(M^2) + (M^2_{l,h} - M^2) I_{\rm log}(M^2)
\nn \\
&& \hspace{1.0cm}+ \, \frac{1}{(4\pi)^2} \left[M^2_{l,h} - M^2 
- M^2_{l,h} \ln \left(\frac{M^2_{l,h}}{M^2}\right)\right].
\label{scale-rel}
\eea

We note that to arrive at these results, no momentum shift was made in 
the divergent integrals; if one had shifted the momenta without change in the integration limits, 
one would have missed the term proportional to $A_{\mu\nu}(M^2)$. We also note that the divergences 
in each of the integrals in the first equality of Eq.~(\ref{Z0-def}) cancel and the final result is
finite. Therefore, there is no need for a regulator in $Z_0$. This feature,
that one can remove the regulator in finite integrals, plays a very important role when considering high
temperatures and densities in the NJL model~\cite{{Farias:2006cs},{Farias:2005cr},
{Farias:2007zz},{Farias:2016let}}. It will play an important role also in the phenomenology of mesons with heavy 
quarks, as we discuss in the next section. 
 
We stress that Eq.~(\ref{EE-fin}) is an exact result and no approximations were made in 
the derivation from Eq.~(\ref{EE-tr}). Moreover, as already mentioned, no momentum 
shifts were made in obtaining the symmetry-violating and divergent integrals. This is 
important, as a momentum shift in a divergent integral is a delicate process and in many instances 
is the source of symmetry violation. Note that whatever choice made for $\eta_\pm$ unavoidably implies 
translation symmetry breaking, {\em unless} the regularization scheme leads to 
$A_{\mu\nu}(M^2) = 0$. Momentum shifts were made only in the {\em finite integral} 
$Z_0(M^2_l,M^2_h,P^2;M^2)$; it can be  integrated without imposing an ultraviolet cutoff. 
The integral develops an imaginary part that reflects the possibility of meson decay into a 
quark-antiquark pair when $P^2 < - (M_l+M_h)^2$; this is an unphysical feature that afflicts 
CI models. However, there is no difficulty in introducing an infrared cutoff~\cite{Ebert:1996vx} in 
this and other finite integrals to avoid unphysical quark-antiquark thresholds in BS amplitudes. This
will be implemented in Sec.~\ref{secIV}.

The expressions for the remaining kernels ${\cal K}^{EF}_{\rm PS}$, ${\cal K}^{FE}_{\rm PS}$, 
and ${\cal K}^{FF}_{\rm PS}$ contain the same terms as in Eq.~\eqref{EE-fin} and an additional symmetry 
violation term{\textemdash}they are presented in Appendix~\ref{BS-kernels}. We also note that one can use the gap 
equation, Eq.~(\ref{gapMf}), to express the quadratically divergent integral, $I_{\rm quad}$, in terms of 
the constituent-quark mass, $M_f$, as
\begin{eqnarray}
\label{quaddiver}
I_{\rm quad}(M^2_f) = \frac{3}{16} \left(m^f_{\rm G}\right)^2 \, 
\frac{\left(M_f - m_f\right)}{M_f} \,.
\label{Iquad-gap}
\end{eqnarray}
We will make use of this equation further ahead.

Let us next examine how choices of $\eta_{\pm}$ lead to violation of the WGT
identity in Eq.~\eqref{eq:avWGT} in case of CI.  In the chiral limit, $m^h = m^l =0$,
combining Eq.~\eqref{eq:avWGT} with Eq.~\eqref{eq:BSEps}, straightforward manipulations 
lead to two equations (for the light-light case):
\begin{eqnarray}
\label{WGT-M}
\hspace{-0.5cm}0 &=& M_l - \frac{8}{3} \left(\frac{1}{m^{l}_{\rm G}}\right)^2 
\int^\Lambda_q \left( \frac{M_l}{q^2_+ + M^2_l} 
+ \frac{M_l}{q^2_- + M^2_l} \right),  \\[0.5true cm] 
\hspace{-0.5cm}0 &=& \int^\Lambda_q 
\left(\frac{q_+\cdot P}{q^2_+ + M^2_l}  - \frac{q_-\cdot P}{q^2_- + M^2_l}\right) .
\label{WGT-gamma}
\end{eqnarray}
Using Eq.~(\ref{Iqua-pm}) in Eq.~\eqref{WGT-M}, one obtains,
\begin{eqnarray}
0 &=& M_l - \frac{16}{3} \left(\frac{1}{m^{l}_{\rm G}}\right)^2 \Bigl[ M_l I_{\rm quad}(M^2_l)  
       \nonumber \\[0.25true cm] 
  && + \,  M_l \left(\eta^2_+ + \eta^2_-\right) A_{\mu\nu}(M^2)P_\mu P_\nu \Bigr].
\label{WGT-gap}
\end{eqnarray}
The last term, being proportional to $\eta^2_+$ and $\eta^2_-$ are symmetry-violating. The subtractions 
in Eq.~(\ref{WGT-gamma}) lead to new symmetry violating terms, in addition to terms proportional to
$A_{\mu\nu}(M^2)$:
\begin{eqnarray}
\int^\Lambda_q \frac{q_\mu }{q^2_\pm + M^2_{l}} &=&  \mp  \eta_\pm P_\mu \, I_{\rm quad}(M^2_{l}) 
\mp \eta_\pm P_\alpha B_{\alpha\mu}(M^2) \nn\\[0.2true cm]
&& \pm  \eta_\pm \left(\eta^2_{\pm}P^2+M^2-M^2_{l}\right)P_\alpha A_{\alpha\mu}(M^2) \nn \\[0.2true cm]
&& \mp   \frac{1}{2}\eta^3_\pm P_{\alpha} A_{\alpha \beta}(M^2) \left(P_\mu P_\beta  
+ P^2 \delta_{\beta \mu}\right) \nn \\[0.2true cm]
&& \mp   \frac{1}{3} \eta^3_{\pm} P_\alpha P_\beta P_\rho\,C_{\alpha\beta\rho\mu}(M^2) ,
\label{WGT-ABC}
\end{eqnarray}
where $B_{\mu\nu}(M^2)$ and $C_{\mu\nu\rho\sigma}(M^2)$ are new tensor structures,
\begin{eqnarray}
B_{\mu\nu}(M^2)  & = &  \int^\Lambda_q   \frac{2q_\mu q_\nu - (q^2+M^2)\delta_{\mu\nu}}{(q^2+M^2)^2}, 
\label{B_munu}
\\[0.3true cm]
C_{\mu\nu\rho\sigma}(M^2)  & = & \int^\Lambda_q   \frac{ c_{\mu \nu \rho \sigma}(q,M^2)}{(q^2+M^2)^4}  ,
\label{C_munurhosigma}
\end{eqnarray}
with
\begin{eqnarray}
c_{\mu\nu\rho\sigma}(q^2,M^2) &= & 24 q_\mu q_\nu q_\rho q_\sigma  - \, 4 (q^2+M^2)  \nonumber \\[0.25true cm]
& \times & ( \delta_{\mu\nu} q_\rho q_\sigma  + \mathrm{perm.} \; \nu\sigma\rho) .
\label{cmunusr}
\end{eqnarray}
The $B_{\mu\nu}(M^2)$ and $C_{\mu\nu\rho\sigma}(M^2)$ integrals are the Euclidean space 
counterparts of the Minkowski space integrals $\nabla_{\mu\nu}(M^2)$ and 
$\square_{\mu\nu\rho\sigma}(M^2)$ of Refs.~\cite{Batt-thesis,Battistel:2008fd,Battistel:2013cja}. 
Using Eq.~\eqref{WGT-ABC} in Eq.~\eqref{WGT-gamma}, we obtain,
\begin{eqnarray}
0 &=& \int^\Lambda_q  \left(\frac{q_+\cdot P}{q^2_+ + M^2_l}  - \frac{q_-\cdot P}{q^2_- + M^2_l}\right) 
\nn\\[0.3true cm]
& = &  \textrm{terms proportional to} \;\; \eta_\pm  \left(A_{\mu \nu}, B_{\mu \nu},
C_{\mu\nu\rho\sigma}\right) .\hspace{0.5cm}
\label{int-ABC}
\end{eqnarray}
We observe that for arbitrary momentum routing in the loop
integrals, the subtraction scheme allows to systematically identify symmetry violating
terms; they are proportional to the integrals $A_{\mu\nu}$, $B_{\mu\nu}$ and $C_{\mu\nu\rho\sigma}$  
in Eqs.~(\ref{A_munu}), (\ref{B_munu}) and (\ref{C_munurhosigma}). A consistent 
regularization scheme must make the integrals vanish automatically. Otherwise, the vanishing of the
integrals must be imposed; in doing so, the regularization scheme becomes a central part of the model. 
Dimensional regularization and Pauli-Villars regularization are examples of schemes that lead to 
$A_{\mu\nu}=0$, $B_{\mu\nu} = 0$, and $C_{\mu\nu\rho\sigma} = 0$. Removing the symmetry-violating terms, Eq.~(\ref{WGT-gap}) becomes nothing else than the gap equation of Eq.~(\ref{gapMf}).  

Let us make contact with the GBCR scheme~\cite{GutierrezGuerrero:2010md}. For a proper comparison,
we need to take $\eta_+ = 1$ and $\eta_- = 0$, the choice made in that reference. For this choice,  
Eq.~(\ref{WGT-gamma}) becomes 
\begin{eqnarray}
\int^\Lambda_q 
\left(\frac{q_+\cdot P}{q^2_+ + M^2_l}  - \frac{q\cdot P}{q^2 + M^2_l}\right) = 0,
\label{WGT-MM}
\end{eqnarray}
which is Eq.~(15) of Ref.~\cite{GutierrezGuerrero:2010md}. Using Eq.~(\ref{WGT-ABC}) 
for this integral, one obtains: 
\begin{eqnarray}
\hspace{-0.25cm}0 &=& \int^\Lambda_q 
\left(\frac{q_+\cdot P}{q^2_+ + M^2}  - \frac{q\cdot P}{q^2 + M^2}\right) 
\nonumber \\[0.2true cm]
&=& P_\mu \bigl[ B_{\mu \nu}(M^2) + \frac{1}{3} C_{\mu\nu\rho\sigma}(M^2) P_\rho P_\sigma 
\nonumber \\[0.2true cm]
&& + \  \frac{1}{3}P_\mu A_{\rho \nu }(M^2) P_\rho 
- \frac{4}{3}P^2 A_{\mu\nu}(M^2) \bigr] P_\nu ,
\end{eqnarray}
that is, the $M_l$ dependence cancels and the remaining divergent terms depend on $M$ only:
\begin{eqnarray}
B_{\mu \nu}(M^2) &+& \frac{1}{3} C_{\mu\nu\rho\sigma}(M^2) P_\rho P_\sigma 
\nonumber \\[0.2true cm]
&& \hspace{-0.75cm} + \, \frac{1}{3} P_\mu A_{\rho \nu }(M^2) P_\rho 
-\frac{4}{3}P^2 A_{\mu\nu}(M^2) = 0.
\end{eqnarray}
In the chiral limit, this is equal to 
\begin{equation}
B_{\mu\nu}(M^2) = 0 = \delta_{\mu\nu} \int_\Lambda \frac{d^4 q}{(2\pi)^4}
\frac{\frac 1 2 q^2 + M^2}{(q^2 + M^2)^2},
\end{equation}
which is Eq.~(17) of Ref.~\cite{GutierrezGuerrero:2010md}. This result makes
it clear that our scheme, besides being in agreement with Ref.~\cite{GutierrezGuerrero:2010md}
for the particular choice of $\eta_\pm$, it is also more general as is valid 
for {\em arbitrary} values of $\eta_\pm$. This is important when $M_h \gg M_l$, as results 
are very sensitive to the momentum partitioning between the heavy and light quarks in a bound
state~\cite{Gomez-Rocha:2016cji}; for $u$ and $s$ quarks, it has been shown~\cite{Maris-Roberts} 
that results for $\pi$ and $K$ observables are not very sensitive to this momentum partitioning.

Here, we take the opportunity to comment on the role played by the subtraction mass~$M$. As mentioned 
previously, it shares similarities with the mass scale~$\mu$ that appears in dimensional regularization. 
Indeed, suppose one uses dimensional regularization in the model. Besides removing automatically 
all potential symmetry-violating terms, dimensional regularization introduces an arbitrary mass 
scale $\mu$ that comes with the replacement $D = 4 \rightarrow D = 4-2\epsilon$ in the integrals.
On the other hand, in the subtraction scheme one introduces the arbitrary mass $M$ and prescribe
that symmetry-violating terms vanish. In addition, one also obtains divergent integrals depending on 
$M$, like $I_{\rm log}(M^2)$ and $I_{\rm quad}(M^2)$ of Eqs.~(\ref{Ilog}) and (\ref{Iquad}), 
that need a regularization that cannot be removed because the model is nonrenormalizable. 
Applying dimensional regularization in those integrals, one introduces two arbitrary 
dimensionful parameters, $M$ and $\mu$: 
\bea
I_{\rm log}(M^2) &=& (\mu)^{2\epsilon}\int \frac{d^Dq}{(2\pi)^D} \frac{1}{(q^2 + M^2)^2}  
\nn \\[0.25true cm]
&=& \frac{1}{(4\pi)^2}\left[\frac{1}{\epsilon} - \ln\left(\frac{M^2}{{\bar\mu}^2}\right)\right],
\eea
where ${\bar \mu}^2 = 4\pi e^{-\gamma} \mu^2$. Choosing $\bar\mu = M$, the integral is 
scale independent; likewise for $I_{\rm quad}(M^2)$, it depends only on $M$. Alternatively,
there is no need for evaluating these integrals with an explicit regulator, since $M$ can be set
by fitting the parameters of the model to physical quantities~\cite{{Battistel:2003gn},{Battistel:2008fd},
{Battistel:2013cja}}. Through relations like that in Eq.~(\ref{scale-rel}), one obtains
the divergent integrals at any other scale in terms of the
physical quantities fitted at the scale $M$. As said, we do not trail this interesting possibility 
in this paper and simply keep $M$ arbitrary and use a second scale that is an ultraviolet cutoff. 
As we shall discuss in the next section, our results are essentially independent of $M$; small
differences in the results caused by using different values can be absorbed by refitting parameters.

The masses of the mesons are obtained from Eq.~\eqref{MHSE_mps}; the equation defines an eigenvalue 
problem with solutions for the masses $P^2 = -m^2_{\rm PS}$, namely:
\bea
\lambda(P^2)\, \Gamma_{\rm PS}(P^2) = { \cal K}(P^2)\, \Gamma_{\rm PS}(P^2),
\label{eigen}
\eea
where ${\cal K}(P^2)$ is the $2\times2$ matrix defined in Eq.~\eqref{MHSE_mps}, and $\lambda(-m^2_{\rm PS})=1$. 
The derivation of the subtracted kernels is detailed in Appendix~\ref{BS-kernels}. 

The weak decay constant of the PS meson, $f_{\rm PS}$, can be extracted from:
\begin{eqnarray}
P_\mu f_{\rm PS} &=& \langle 0|\bar\psi_l(0)\gamma_\mu\gamma_5\psi_h(0)|\phi(P)\rangle \nn \\
&=& N_c\int_q {\rm Tr}\left[\gamma_5\,\gamma_\mu S_l(q_+)\Gamma^{lh}_{\rm PS}(P)S_h(q_-)
\right],
\label{def-fps}
\end{eqnarray}
where the trace is over Dirac indices. Inserting Eq.~\eqref{contactBSA} into Eq.~\eqref{def-fps} and 
evaluating the trace leads to,  
\begin{equation}
\hspace{-0.1cm}f_{\rm PS} = \frac{N_c}{4M_{lh}}\left[E_{\rm PS}(P){\cal K}^{FE}_{\rm PS}
           +  F_{\rm PS}(P){\cal K}^{FF}_{\rm PS}\right]_{P^2=-m^2_{\rm PS}},
\label{fps}
\end{equation}
where, we recall, $M_{lh} = M_lM_h/(M_l+M_h)$. The eigenvalue equation in Eq.~(\ref{eigen}) does not 
fix the amplitudes $E_{\rm PS}(P)$ and $F_{\rm PS}(P)$ separately; for that, one needs a normalization 
condition, which we take as,
\bea
2 P_\mu &=& N_c \int 
\frac{d^4q}{(2\pi)^4} {\rm Tr} \Biggl[\Gamma^{hl}_{\rm PS}(-P) \frac{\partial S_l(q_+) }{\partial P_\mu} 
\Gamma^{hl}_{\rm PS}(P)
S_h(q_-) \nn \\
&& + \, \Gamma^{hl}_{\rm PS}(-P) S_l(q_+) \Gamma^{hl}_{\rm PS}(P)\frac{\partial S_h(q_-)}{\partial P_\mu}
\Biggr], 
\label{norm-ps} 
\eea
evaluated at $P^2=-m^2_{\rm PS}$. We note that with this convention, the experimental value of the pion decay 
constant is $f_\pi \simeq 130$~MeV~\cite{PDG}. 

To close this section, we comment on the regularization procedure. On computing the masses and the decay 
constants of the pseudoscalar mesons, the first step is to obtain the constituent quark masses, $M_f$,  
from the gap equation, Eq.~\eqref{gapequ}, and then solve the eigenvalue problem, 
Eq.~\eqref{eigen}. Both equations contain the ultraviolet divergent integrals $I_{\rm quad}$ and
$I_{\rm log}$ that need regularization and, in addition, Eq.~\eqref{eigen} contains also the finite integral
$Z_0$-function (and the $Z_1$ that appears in the other kernels). 
As briefly discussed in Sec.~\ref{secIII}, $Z_0$ develops thresholds as it can acquire an imaginary 
part when $P^2 < - (M_l+M_h)^2$. We avoid the possibility of unphysical thresholds in hadron decay 
into quarks by eliminating this branch cut with an infrared cutoff. Both ultraviolet and infrared 
cutoffs can be introduced using a proper-time regularization~\cite{Ebert:1996vx} as follows. All 
integrals, finite and divergent, contain integrands of the form $1/(q^2 + a^2)^n$, where $n=1, 2$ 
and $a^2$ is independent of $q$. The imaginary part in $Z_0$ happens when $q^2 \rightarrow - a^2$; 
to avoid this to happen, one then rewrites the integrands as 
\begin{eqnarray}
\frac{1}{(q^2 + a^2)^n} &=& \frac{1}{(n-1)!} \int^\infty_0 d\tau \, \tau^{n-1} \,
e^{-\tau(q^2 + a^2)} \nn \\[0.2true cm]
&\rightarrow& \frac{1}{(n-1)!} \int^{\tau_{\rm ir}}_{\tau_{\rm uv}} d\tau \, \tau^{n-1} \,
e^{-\tau(q^2 + a^2)} ,
\end{eqnarray}
with $\tau_{\rm uv} = 1/\Lambda^2_{\rm uv}$ and $\tau_{\rm ir} = 1/\Lambda^2_{\rm uv}$, where
$\Lambda_{\rm uv}$ and $\Lambda_{\rm ir}$ are the ultraviolet and infrared cutoffs respectively.
For the $Z_0$ integral, $\tau_{\rm uv}$ is not needed, as the integral is finite; therefore 
\begin{equation}
\frac{1}{(q^2 + a^2)^2} \rightarrow \frac{1 - \left[1 + \left(q^2 + a^2\right)\tau_{\rm ir}\right] e^{-\tau_{\rm ir} 
(q^2 + a^2)}}{(q^2 + a^2)^2},
\end{equation}
which is finite when $q^2 \rightarrow - a^2$, as can be verified very easily. The final result for
$Z_0$ can be written as 
\begin{eqnarray}
Z_0(M^2_l,M^2_h,P^2;M^2)&=&\frac{1}{(4\pi)^2}\int^1_0dz 
\int^{\tau^2_{\rm ir}}_{0}\frac{d\tau}{\tau} e^{-\tau M^2} 
\nn\\[0.1true cm]
&& \times \, \left( 1 - e^{-\tau [H(z) - M^2]}\right),
\end{eqnarray}
which is clearly finite when $\tau \rightarrow 0$. Unphysical thresholds in the function $Z_1$
are eliminated using its relation to $Z_0$, given in Eq.~(\ref{Z1-def}).

Although $\tau_{\rm ir}$ is not needed for $I_{\rm quad}$ and $I_{\rm log}$,  we keep it in those integrals 
to compare results with Ref.~\cite{GutierrezGuerrero:2010md}. Therefore
\begin{equation}
I_{\rm quad}(M^2_f)  =  \frac{1}{(4\pi)^2} \int^{\tau^2_{\rm ir}}_{\tau^2_{\rm uv}} 
\frac{d\tau}{\tau^2}\,  e^{- \tau M^2_f} .
 \label{quadraticreg}
\end{equation}
The logarithmically divergent integral $I_{\rm log}(M^2_f)$ can be derived from $I_{\rm quad}(M^2_f)$
by a simple differentiation: 
\begin{eqnarray}
I_{\rm log}(M^2_f) & = &  -\frac{\partial  I_{\rm quad}(M^2_f)}{\partial M^2_f} . 
\end{eqnarray}

%
\section{Numerical Results}
\label{secIV}

We start considering the $f=\{u,d,s\}$ quark flavors and compare results from the CI subtraction 
and GBCR schemes~\cite{GutierrezGuerrero:2010md}. More specifically, we compare results with 
Ref.~\cite{Chen:2012txa} that extended the GBCR CI to the strange flavor sector.
We recall the differences that mark both approaches: in the subtraction scheme the results are 
independent of $\eta_\pm$, while those in Refs.~\cite{{GutierrezGuerrero:2010md},
{Chen:2012txa}} are for $\eta_+ = 1$ and $\eta_- =0$. The finite integrals $Z_0$ and $Z_1$ in the
BS amplitudes that appear in the subtraction scheme can be integrated without imposing ultraviolet or 
infrared cutoffs, though we removed the thresholds of these functions by means of infrared cutoff as 
discussed in the previous section. As shown in Sec.~\ref{secIII}, for $\eta_+ = 1$ and $\eta_- =0$ and neglecting 
symmetry violating terms and using the same ultraviolet and infrared cutoffs, both approaches lead to 
identical results.

The free parameters are the quark masses $m_u$, $m_d$ and $m_s$, the coupling strength 
$\alpha_{\rm IR}$, the gluon-mass scale $m_g$, and the cutoffs $\Lambda_{\rm ir}$ 
and $\Lambda_{\rm uv}$.  For a proper comparison, we use the parameter set of Ref.~\cite{Chen:2012txa}:
$m_u=m_d=m = 7$~MeV, $m_s = 170$~MeV, $\alpha_{\rm IR} = 0.93\pi$, $m_g = 800$~MeV, $\Lambda_{\rm ir} = 240$~MeV, 
and $\Lambda_{\rm uv} = 905$~MeV. The ultraviolet cutoff is used in the divergent integrals
$I_{\rm log}$ and $I_{\rm quad}$ only. These parameters give for the constituent masses, solutions 
of the gap equation in Eq.~(\ref{gapMf}):  $M_u = M_d = 367$~MeV and $M_s = 523$~MeV.

In Tab.~\ref{tab:comp} we present the BS amplitudes $(E,F)_{\rm PS}$ and masses and decay constants 
$(m,f)_{\rm PS}$ of $\pi$ and $K$ from both approaches. Here, we use the natural value for the subtraction 
mass $M$ in the calculation of for both $\pi$ and $K$ properties, namely the light constituent quark mass, 
$M = M_u$; choosing $M = M_s$, or any other value between $M_u$ and $M_s$, does not change the results
for the properties of $\pi$ and $K$, as expected. Table~\ref{tab:comp} informs that the results in both 
approaches compare very well, and are in good agreement with experimental data. The particular choice of 
momentum partitioning and use or not of an ultraviolet cutoff in finite integrals does not affect the 
results in this case. While the first conclusion is consistent with the findings in 
Ref.~\cite{Maris-Roberts} in a finite-range RL model, the second means that the momentum integrands 
that give rise to the finite functions $Z_0$ and $Z_1$ have integrands well concentrated within the 
momentum range extending from $\Lambda_{\rm ir}$ to $\Lambda_{\rm uv}$.      

\begin{table}[h]
\caption{\label{tab:comp} BS amplitudes $(E,F)_{\rm PS}$ and masses and
decay constants $(m,f)_{\rm PS}$ of $\pi$ and $K$ from the CI subtraction
scheme and CI GBCR of Ref.~\cite{Chen:2012txa}.
The amplitudes are dimensionless and masses and decay constants are listed in MeV.}
\begin{ruledtabular}
\begin{tabular}{lcccccc}
& $(E,F)_\pi$ & $(E,F)_K$  & $(m,f)_\pi$ & $(m,f)_K$ \\[0.2true cm]
\cline{2-5} \\[-0.2true cm]
Data~\cite{PDG}                 & ---          & ---           & (139, 131) & (494, 156) \\ [0.1true cm]
CI-subtr                        & (5.10, 0.67) & (5.64, 0.85)  & (139, 144) & (502, 153)  \\
GBCR~\cite{Chen:2012txa}      & (5.09, 0.68) & (5.40, 0.83)  & (140, 141) & (500, 156)  \\            
\end{tabular}
\end{ruledtabular}
\end{table}

Next, we include the charm quark and calculate properties of heavy-light mesons. The issue on 
the necessity for a flavor dependence of the effective coupling is exposed: using the same parameters as in 
the light quark sector and adjusting the charm quark mass to fit the mass of the $D-$meson, the decay constant 
$f_D$ turns out smaller than $f_K$. The same feature is also present in Ref.~\cite{blaschke} within the 
NJL model. Like in the extension of the NJL model made in Refs.~\cite{{Casalbuoni:2003cs},{Farias:2006cs}} to high
temperatures and densities, and also in the extension of the GBCR CI model in Refs.~\cite{{Bedolla:2015mpa},
{Bedolla:2016yxq}} to charmonium, one can readjust the effective coupling $m^h_{\rm G}$ and the 
ultraviolet cutoff $\Lambda_{\rm uv}$ in our model. In the light quark sector, we used $m_g = 800$~MeV and 
$\alpha_{\rm IR} = 0.93\pi$, which imply $m^{l}_{\rm G} = 132$~MeV. We set $m_c = 1454$~MeV, and reset
$\Lambda_{\rm uv} = 1290$~MeV and $m^h_{\rm G} = 3.5 \, m^l_{\rm G} = 462$~MeV{\textemdash}see 
Eq.~\ref{eq:Kcontact}. We stress that the new value 
of $\Lambda_{\rm uv}$ is used in the gap and BS equation involving the charm quark only. Regarding the 
subtraction mass $M$, we use for it the light quark mass; more on this at the end of this section. From 
the gap equation, we obtain $M_s = 529$~MeV and $M_c = 1490$~MeV, which shows that there is very little 
dressing of the charm quark. 

\begin{table}[h]
\caption{\label{tab:ps-mes} Masses and decay constants of the pseudoscalar  
mesons (in MeV). Parameters of the CI subtraction model are given in the main text. 
NST1, NST2, REBM1, REBM2, HGKL1 and HGKL2 are results from representative studies employing 
finite-range interactions treated within the RL framework of the DS and BS equations in QCD.}
\begin{ruledtabular}
\begin{tabular}{lcccc}
& \multicolumn{4}{c}{$(m, f)_{\rm PS}$}  \\[0.1true cm] 
\cline{2-5} \\[-0.25true cm]
& $\pi$ & $K$ & $D$ & $D_s$ \\[0.1true cm]
\hline \\[-0.2true cm]
Data~\cite{PDG}                 & (139, 131) & (494, 156) & (1864, 212) & (1968, 249) \\ 
CI-subtr                        & (139, 143) & (494, 153) & (1869, 207) & (1977, 240) \\
GBCR~\cite{Chen:2012qr}       & (140, 141)  & (500, 156) &     ---     &    ---       \\ 
                                & {}         & {}         & {}          & {}           \\[-0.25true cm]
NJL~\cite{blaschke}             & (135, 131)  & (498, 135)  & (1869, 113) & --- \\
NST1~\cite{Nguyen:2010yh}       & (138, 131)  & (497, 155)  & (1850, 222) & (1970, 197)  \\
NST2~\cite{Nguyen:2010yh}       &    ---     &    ---       & (1880, 260) & (1900, 275)  \\  
REBM1~\cite{Rojas:2014aka}      & (138, 139)  & (493, 164)  & (2115, 204) & (2130, 249) \\ 
REBM2~\cite{Rojas:2014aka}      & (153, 189)  & (541, 214)  & (2255, 281) & (2284, 320) \\
HGKL1~\cite{Hilger:2017jti}     & (137, 133)  & (492, 155)  & (1868, 323) & (1872, 269) \\
HGKL2~\cite{Hilger:2017jti}     & (137, 128)  & (489, 150)  & (1869, 960) & (1802, 295)  
\end{tabular}
\end{ruledtabular}
\end{table}

Table.~\ref{tab:ps-mes} presents our results for the masses and decays constants of
pseudoscalar mesons. We have also readjusted the strange quark mass to $m_s = 165$~MeV to obtain 
a perfect fit to $m_K$. For orientation and comparison, we have listed in the table the 
results from the GBCR CI and from several representative studies employing finite-range interactions 
treated within the RL framework of the DS and BS equations in QCD. The latter are identified in the 
table by NST1 and NST2 from Ref.~\cite{Nguyen:2010yh}, REBM1 and REBM2 from Ref.~\cite{Rojas:2014aka} 
and HGKL1 and HGKL2 from Ref.~\cite{Hilger:2017jti}. The first observation one can draw from 
Tab.~\ref{tab:ps-mes} is that, once the parameters of CI subtraction model are fixed to obtain an 
almost perfect agreement with experiment for the $m_{\rm PS}$ masses, the model predicts for the decay 
constants, $f_\pi < f_K < f_D < f_{D_s}$, a pattern that is corroborated by realistic DS-BS and 
lattice-QCD calculations. Moreover, the individual values of $f_{\rm PS}$ 
are in very good agreement with the experimental and lattice data collected by the PDG~\cite{PDG}, 
the discrepancies being at the level of a few percent. It is also remarkable that the CI subtraction 
model gives a much better description for $(m, f)_{\rm PS}$ than finite-range models within the RL 
framework, in which the discrepancies with the PDG values for the decay constants can reach up to 40\%
in some cases. 

Regarding the very small value of $f_D$ in the NJL model, one may reasonably object that it is due to 
the inadequate use of parameters adjusted to the light mesons. Although it is 
true that one can remedy this situation adjusting the coupling and cutoff in the NJL model, one still 
faces the problem that the results become unacceptably dependent on the choices for~$\eta_\pm$. This means
that for any new choice of $\eta_\pm$, new values for the parameters of the model are required
to fit observables; the situation is even more dramatic, in that no solutions for the BS equation 
can be found for some choices of $\eta_\pm$. The subtraction scheme solves this problem.

We have also calculated masses and decay constants of vector mesons with the same parameter set
used for the pseudoscalar mesons; the results are listed in Tab.~\ref{tab:v-mes}. There is fairly good 
experimental information on their masses but not on their decay constants. The values quoted in 
Tab.~\ref{tab:v-mes} for the masses are taken from PDG~\cite{PDG} and those for the decay 
constantes of $\rho$ and $K^*$ are extracted indirectly~\cite{Maris:1999nt} from $\tau$~decays, 
whereas those for $D^*$ and $D^*_s$ are from a recent lattice calculation~\cite{Becirevic:2012ti}. 
The CI predictions for the masses, $m_{\rm V}$, agree very well with experiment 
but less well for $f_{\rm V}$, as there is a wrong ordering between $f_\rho$ and $f_{K^*}$ and between $f_{D^*}$ and 
$f_{D^*_s}$. On the other hand, inspection of Tab.~\ref{tab:v-mes} reveals that finite-range 
RL models also have difficulties in the vector sector; for instance, $f_{D^*_s} < f_{K^*}$ in 
all of those models. In this connection, also calculated charmonium properties: 
in the same vein of Ref.~\cite{Bedolla:2015mpa}, by a further increase of $m^h_G$ and $\Lambda_{\rm uv}$, 
the CI subtraction model predicts similar results to those of the GBCR 
model, including the wrong ordering between the vector and pseudoscalar decay constants, 
$f_{J/\Psi} < f_{\eta_c}$. This is an additional indication that the vector channels pose a challenge 
for CI models, possibly reflecting in this context the limitations of the RL framework exposed in 
finite-range models~\cite{{Maris:2005tt},{Bashir:2012fs},{Binosi:2016rxz},{Eichmann:2016yit}}.

\begin{table}[htb]
\caption{\label{tab:v-mes} Masses and decay constants of vector mesons (in MeV).
Parameters of the CI subtraction model are the same as in Tab.~\ref{tab:ps-mes}. The entries 
for Data are explained in the text.} 
\begin{ruledtabular}
\begin{tabular}{lcccc}
& \multicolumn{4}{c}{$(m, f)_V$}  \\[0.1true cm] 
\cline{2-5} \\[-0.25true cm]
& $\rho$ & $K^*$ & $D^*$ & $D^*_s$ \\[0.1true cm]
\cline{2-5} \\[-0.1true cm]
Data                            & (775, 212)  & (892, 225)  & (2010, 278) & (2112, 322)  \\
CI-subt                         & (776, 205)  & (881, 195)  & (2011, 281) & (2098, 276)  \\
GBCR~\cite{Chen:2012txa}        & (930,\;---\;) & (1030,\;---\;) &     ---    &    ---       \\ 
GBCR~\cite{Roberts:2011wy}      & (928,182) & (\;---\;,\;---\;) &     ---    &    ---       \\ 
                                & {}    & {}    & {} & {}                            \\[-0.25true cm]
NST1~\cite{Nguyen:2010yh}       & (742, 207)  & (936, 241) & (2040, 160) & (2170, 180) \\
HGKL1~\cite{Hilger:2017jti}     & (758, 219)  & (946, 247) &    ---      & (2175, 178) \\
HGKL2~\cite{Hilger:2017jti}     & (725, 203)  & (919, 237) &    ---      &    ---  
\end{tabular}
\end{ruledtabular}
\end{table}

%
\section{Summary and Conclusion}
\label{secV}

We examined the CI model introduced in Ref.~\cite{GutierrezGuerrero:2010md}
within the perspective of a regularization scheme that allows to separate symmetry-violating 
parts in BS amplitudes in a choice independent of the momentum partition
between the quark and antiquark in the bound state. In doing so, spacetime-translation symmetry
and the WGT identities reflecting global symmetries of the model are 
preserved by the regularization. Symmetry-offending parts of the amplitudes, the integrals 
$A_{\mu\nu}$, $B_{\mu\nu}$, $C_{\mu\nu}$ and $D_{\mu\nu}$, can be neatly separated. In general, 
a cutoff regularization scheme leads to nonzero values for the symmetry-violating integrals, 
while dimensional regularization leads to the vanishing of the symmetry-offending integrals. In 
a nonrenormalizable model, like the contact-interaction model discussed here, the vanishing of 
$A_{\mu\nu}$, $B_{\mu\nu}$ and $C_{\mu\nu}$ must be imposed in an {\em ad hoc} manner and
te imposition becomes an integral part of the model. 

In order to check the subtraction scheme we have studied properties of the heavy-light $D$, 
$D_s$, $D^*$ and $D_s$ mesons, for which symmetry offending terms in a CI model have catastrophic 
consequences, as well as of light mesons $\pi$, $K$, $\rho$ and $K^*$. We have shown that using the subtraction 
scheme we can obtain values for the masses of all of these mesons that agree well with experiment.
In addition, we can obtain the correct trend of the weak decay constants of the pseudoscalar
mesons, $f_\pi < f_K < f_D < f_{D_s}$, and their individual values are also in good agreement 
with experiment and lattice QCD simulations. The values of the decay constants of the $\rho$ and 
$D^*$ agree very well with experiment and lattice QCD simulations, respectively; but the model
predicts the wrong orderings $f_\rho < f_K$ and $f_{D^*} < f_{D^*_s}$. These difficulties in the
vector sector might not be related to intrinsic deficiencies of a contact interaction, as similar
deficiencies have been observed in RL calculations using the DS-BS equations employing sophisticated 
finite-range interaction kernels.

We reiterate that the CI subtraction scheme developed here is not meant to be a substitute 
for a full-fledged, QCD-based DS-BS framework. Rather, its aspiration is to be a simpler tool 
that preserves spacetime symmetries and global symmetries of QCD. We strongly believe that 
a simpler tool that respects such basic symmetries, and at the same time, is able to give 
comparable results of more sophisticated and intricate approaches, acquires the necessary 
credentials to be used to explore and provide useful insight into complex problems of current 
interest, like multihadron molecules and hadronic matter at finite temperature and density. 

As a next application of the subtraction scheme we mention the study of strong couplings of charmed
hadrons with light hadrons, in particular the couplings of $D-$mesons to nucleons. Such couplings 
are relevant in studies of the $D-$nucleon interaction at low energies~\cite{Haidenbauer:2007jq,
Haidenbauer:2008ff,Haidenbauer:2010ch,Fontoura:2012mz}, $D-$mesic nuclei~\cite{{Tsushima:1998ru},
{GarciaRecio:2010vt},{GarciaRecio:2011xt}}, $J/\psi$ binding to nuclei~\cite{{Ko:2000jx},{Krein:2010vp},
{Tsushima:2011kh}}, among others. There is no experimental information about the $D-$nucleon interaction, 
most of the knowledge on 
this interaction comes from calculations using hadronic Lagrangians motivated by SU(4) 
extensions of light-flavor chiral Lagrangians. Since the symmetry is badly broken by the widely
different values of the quark masses, it is important to check the validity of SU(4) symmetry
in the couplings not only in vacuum~\cite{{Krein:2012lra},{ElBennich:2011py},{Krein:2014vma},{El-Bennich:2016bno},
{Ballon-Bayona:2017bwk}}, but at finite temperature and density as well. Another future application
is to study charm meson diffusion in mesonic matter. Here it is necessary to compute for example the 
temperature dependence of masses of $D^+$ $(D^-)$ mesons and also the light mesons $\pi$, $K$ and 
$\eta$ \cite{GhoshSernaKrein}. With these inputs, one is able to compute the drag and diffusion of $D^+$ 
meson in hadronic medium.

\section*{Acknowledgments}
This work of was supported in part by Conselho Nacional de Desenvolvimento  Cient\'{\i}fico e 
Tecnol\'ogico  -- CNPq, Grants nos. 140041/2014-1 (F.E.S.), 458371/2014-9 (B.E.-B), 
305894/2009-9 (G.K.), Funda\c{c}\~ao de Amparo \`a Pesquisa do Estado de  S\~ao Paulo - FAPESP, 
Grant No. 2013/01907-0 (G.K), 2016/03154-7 (B.E.-B.) and by a doctoral scholarships by 
Coordena\c{c}\~ao de Aperfei\c{c}oamento de Pessoal de N\'{\i}vel Superior -- CAPES (F.E.S.).

\appendix

\begin{widetext}

\section{Bethe-Salpeter kernels}
\label{BS-kernels}

Here we collect the expressions of the Bethe-Salpeter kernels for pseudoscalar and
vector and vector channels.

\subsection{Pseudoscalar kernels}

After taking Dirac traces and performing the necessary number of 
subtractions in Eqs.~\eqref{EE}--\eqref{FF} , we arrive at the expressions.
\bea
{\cal K}^{EF}_{\rm PS}  &=&  \frac{P^2} {2M^2_{lh}}{\cal K}^{FE}_{\rm PS} 
= \frac{8P^2}{M_{lh}}\bigg\{\frac{M_h+M_l}{2}I_{\rm log}(M^2)    
+ (M_l-M_h)Z_1(M^2_l,M^2_h,P^2;M^2)  - M_lZ_0(M^2_l,M^2_h,P^2;M^2)\bigg\} \nn\\[0.35true cm]
&& + \, \frac{4A_{\mu\nu}(M^2)}{M_{lh}}(M_l-M_h)P_\mu P_\nu(\eta_+ -\eta_-) ,   \\
{\cal K}^{FF}_{\rm PS}  &=& 8\frac{(M^2_l-M^2_h)}{2P^2} \left[I_{\rm quad}(M^2_h) 
- I_{\rm quad}(M^2_l) \right ] 
-  \frac{8(M_l+M_h)^2[P^2 + (M_l-M_h)^2]}{2P^2} \nn \\
&& \times \, \bigg [ I_{\rm log}(M^2) 
-  Z_0(M^2_l,M^2_h,P^2;M^2)\bigg] + \frac{8}{P^2} \, P_\mu P_\nu D_{\mu\nu}(M^2), 
\label{KFF}
\eea
where $D_{\mu\nu}(M^2)$ is given by
\begin{eqnarray}
\label{Dmunu}
D_{\mu\nu}(M^2) &=& B_{\mu\nu}(M^2)-\frac{(\eta_+ -\eta_-)^2P_\mu P_\alpha}{2}A_{\alpha\nu}(M^2)
-\frac{(\eta_+ -\eta_-)^2P_\nu P_\beta}{2}A_{\beta\mu}(M^2) 
\nn\\[0.35true cm]
&-& \frac{[P^2 (\eta^2_+ -\eta^2_-)+M^2_h-M^2_l]}{2}A_{\mu\nu}(M^2)
+ \frac{(\eta^2_++\eta^2_--\eta_+\eta_-)P_\alpha P_\beta}{3}\bigg[C_{\alpha\beta\mu\nu}(M^2)
\nn\\[0.35true cm]
&+&\delta_{\alpha\beta}A_{\mu\nu}(M^2_h)+\delta_{\alpha\mu}A_{\beta\nu}(M^2)+\delta_{\alpha\nu}A_{\mu\beta}(M^2)\bigg]
- 2P_{\alpha}P_{\beta}[\eta^2_+A_{\alpha\beta}(M^2)+ \eta^2_-A_{\alpha\beta}(M^2)]~.
\end{eqnarray}

For the normalization of the BS amplitudes $E^{lh}_{\rm PS}$ and $F^{lh}_{\rm PS}$, stated by
Eq.~(\ref{norm-ps}) we get 
\begin{eqnarray}
\label{ResulNC}
1 &=&- \frac{3}{2}\bigg\{\left(E^{lh}_{\rm PS}\right)^2 \frac{\partial {\cal K}^{EE}_{\rm PS}}{\partial P^2} 
+ 2 E^{lh}_{\rm PS} F^{lh}_{\rm PS} \left[-\frac{1}{2P^2}{\cal K}^{EF}_{\rm PS}
+\frac{\partial {\cal K}^{EF}_{\rm PS}}{\partial P^2}\right]
+ \frac{\left(F^{lh}_{\rm PS}\right)^2 P^2}{2M^2_{lh}}\frac{\partial{\cal K}^{FF}_{\rm PS} }{\partial P^2}
\bigg\}\bigg\arrowvert_{P^2=-m^2_{\rm PS}},    
\label{normaliz} 
\end{eqnarray}
with the derivatives of the kernels given by the following expressions: 
\begin{eqnarray}
\frac{\partial {\cal K}^{EE}_{\rm PS}}{\partial P^2}&=& - 8 \left[I_{\rm log}(M^2)
-Z_0(M^2_l,M^2_h,P^2;M^2)\right]
- 8 [P^2+(M_l-M_h)^2]Y_1(M^2_l,M^2_h,P^2;M^2) \\[0.25true cm]
\frac{\partial {\cal K}^{EF}_{\rm PS}}{\partial P^2}&=& \frac{{\cal K}^{EF}_{\rm PS}}{P^2}
+ \frac{8P^2}{M_{lh}}\bigl[(M_l-M_h)Y_2(M^2_l,M^2_h,P^2;M^2)
- M_lY_1(M^2_l,M^2_h,P^2;M^2)\bigr], \\[0.25true cm]
\frac{\partial {\cal K}^{FF}_{\rm PS}}{\partial P^2}&=& -\frac{{\cal K}^{FF}_{\rm PS}}{P^2}
- 8 \frac{(M_l+M_h)^2}{2P^2}\bigg\{\bigl[I_{\rm log}(M^2)  - \, Z_0(M^2_l,M^2_h,P^2;M^2)\bigr] \nn \\
&& + \, [P^2+(M_l-M_h)^2] Y_1(M^2_l,M^2_h,P^2;M^2)\bigg\},
\end{eqnarray}
where we have defined
\begin{eqnarray}
Y_1(M^2_l,M^2_h,P^2;M^2)&=&\frac{1}{(4\pi)^2}\int^1_0dzz(1-z)
\left[e^{-H(z)\tau^ 2_{\rm ir}}-1\right], \\[0.25true cm]
Y_2(M^2_l,M^2_h,P^2;M^2)&=& \frac{1}{P2}Z_1(M^2_l,M^2_h,P^2;M^2) 
\nn \\
&& + \, \frac{1}{2P^2}\biggl[Z_0(M^2_l,M^2_h,P^2;M^2) + (P^2+M^2_h-M^2_l)
Y_1(M^2_l,M^2_h,P^2;M^2)\biggr].
\end{eqnarray}
\begin{eqnarray}
\end{eqnarray}
From the eigenvalue equation, one obtains
\beq
\label{Enor}
E^{lh}_{\rm PS}(P) = \frac{{\cal K}^{EF}_{\rm PS}}{\left(\frac{3m^2_G}{4\pi\alpha_{\rm IR}}
- {\cal K}^{EE}_{\rm PS}\right)}\, F^{lh}_{\rm PS}(P). 
\eeq 
Combining this with Eq.~(\ref{normaliz}), the individual amplitudes $E^{lh}_{\rm PS}(P)$ 
and $F^{lh}_{\rm PS}(P)$ are determined.

\section{Vector kernel}

After performing  the necessary number of subtractions in the integrand of Eq.~(\ref{KEE-V}),
one obtains again finite integrals and divergent integrals that are independent of $\eta_\pm$ and
symmetry violating terms:

\bea
{\cal K}^{EE}_{\rm V}(P) &=& \frac{4}{3}\bigg\{
- [P^2+(\Delta M_{lh})^2- 4M_lM_h]\left[I_{\rm log}(M^2)-Z_0(M^2_l,M^2_h,P^2,M^2)\right]
\nn\\[0.2true cm]
&& + \, 2 (P^2+M^2_h-M^2_l)\bigg[Z_0(M^2_l,M^2_h,P^2;M^2) - Z_1(M^2_l,M^2_h,P^2;M^2)
- \frac{1}{2}I_{\rm log}(M^2_l)\bigg] \nn \\
&& + \, 3 \, I_{\rm quad}(M_l) + I_{\rm quad}(M_h) + \left(\eta^2_+ +\eta^2_- \right)A_{\mu\nu}(M^2)
P_\mu P_\nu + \frac{1}{2}\frac{P_\mu P_\nu D_{\mu\nu}(M^2)}{P^2}\bigg\} ,
\eea
where $Z_1(M^2_l,M^2_h,P^2;M^2)$ is another finite integral that can be related~\cite{Batt-thesis}
to $Z_0$ as 
\bea
Z_1(M^2_l,M^2_h,P^2;M^2) &=& \frac{1}{(4\pi)^2} \int^1_0dz\, z \ln \left[\frac{H(z)}{M^2}\right] 
= \frac{1}{2P^2}\bigg\{ M^2_h \left[1 -  \ln\left(\frac{M^2_h}{M^2}\right)\right]
- M^2_l\left[1 - \ln\left(\frac{M^2_l}{M^2}\right)\right] \nn \\
&& + \, (P^2+M^2_h-M^2_l) Z_0(M^2_l,M^2_h,P^2;M^2)\bigg\} .
\label{Z1-def}
\eea 
The term $D_{\mu\nu}(M^2)$ is defined in Eq.~(\ref{Dmunu}).

The vector meson decay constant is given by
\bea
f_{\rm V}=\frac{N_c}{2}E^{lh}_V(P) {\cal K}^{EE}_{\rm V}(P)\bigg|_{P^2=-m^2_V},
\eea
where the BS amplitude is normalized as in the case of pseudoscalar mesons and is 
is fixed as
\bea
\frac{1}{(E^{lh}_V)^2} = N_c \, \frac{\partial {\cal K}^{EE}_{\rm V}(P)}{\partial P^2}\bigg|_{P^2=-m^2_V}~.
\eea

\end{widetext}

%


\begin{thebibliography}{99}    

\bibitem{Manohar:2000dt} 
  A.~V.~Manohar and M.~B.~Wise, 
  \newblock {\em Heavy quark physics} (Cambridge University Press, 
  Cambridge, 2000).
%
\bibitem{Petrov:2016azi} 
  A.~A.~Petrov and A.~E.~Blechman, 
  \newblock {\em Effective Field Theories} 
  (World Scientific, Singapore, 2016).
%
\bibitem{Roberts:1994dr} 
  C.~D.~Roberts and A.~G.~Williams,
  \newblock Prog.\ Part.\ Nucl.\ Phys.\  {\bf 33}, 477 (1994).
%
\bibitem{Alkofer:2000wg} 
  R.~Alkofer and L.~von Smekal,
  \newblock Phys.\ Rept.\  {\bf 353}, 281 (2001).
%
 \bibitem{Cloet:2013jya} 
  I.~C.~Cloet and C.~D.~Roberts,
  \newblock Prog.\ Part.\ Nucl.\ Phys.\  {\bf 77}, 1 (2014).

%
\bibitem{Eichmann:2016yit} 
  G.~Eichmann, H.~Sanchis-Alepuz, R.~Williams, R.~Alkofer and C.~S.~Fischer,
  \newblock Prog.\ Part.\ Nucl.\ Phys.\  {\bf 91}, 1 (2016).
%
\bibitem{Binosi:2016rxz} 
  D.~Binosi, L.~Chang, J.~Papavassiliou, S.~X.~Qin and C.~D.~Roberts,
  \newblock Phys.\ Rev.\ D {\bf 93}, 096010 (2016).
%
\bibitem{Maris:2005tt} 
  P.~Maris and P.~C.~Tandy,
  \newblock Nucl.\ Phys.\ Proc.\ Suppl.\  {\bf 161}, 136 (2006).
%
\bibitem{Nguyen:2010yh} 
  T.~Nguyen, N.~A.~Souchlas and P.~C.~Tandy,
  \newblock AIP Conf.\ Proc.\  {\bf 1361}, 142 (2011)
%
\bibitem{Souchlas:2010boa} 
  N.~Souchlas,
  \newblock J.\ Phys.\ G {\bf 37}, 115001 (2010).
%
\bibitem{Souchlas:2010zz} 
  N.~Souchlas and D. Stratakis ,
  \newblock Phys.\ Rev.\ D {\bf 81}, 114019 (2010).
%
\bibitem{Bashir:2012fs} 
  A.~Bashir, L.~Chang, I.~C.~Clo\"et, B.~El-Bennich, Y.~X.~Liu, C.~D.~Roberts and P.~C.~Tandy,
  \newblock Commun.\ Theor.\ Phys.\  {\bf 58}, 79 (2012).
%
\bibitem{Gomez-Rocha:2014vsa} 
  M.~G\'omez-Rocha, T.~Hilger and A.~Krassnigg,
  \newblock Few Body Syst.\  {\bf 56}, 475 (2015).
%
%
\bibitem{Rojas:2014aka} 
  E.~Rojas, B.~El-Bennich and J.~P.~B.~C.~de Melo,
  \newblock Phys.\ Rev.\ D {\bf 90}, 074025 (2014).
%
\bibitem{Gomez-Rocha:2015qga} 
  M.~Gomez-Rocha, T.~Hilger and A.~Krassnigg,
  \newblock Phys.\ Rev.\ D {\bf 92},  054030 (2015).
%
\bibitem{Gomez-Rocha:2016cji} 
  M.~G\'omez-Rocha, T.~Hilger and A.~Krassnigg,
  \newblock Phys.\ Rev.\ D {\bf 93}, 074010 (2016).
%
\bibitem{Hilger:2017jti} 
  T.~Hilger, M.~Gómez-Rocha, A.~Krassnigg and W.~Lucha,
  arXiv:1702.06262 [hep-ph].
%
\bibitem{Briceno:2015rlt} 
  R.~A.~Brice\~no {\it et al.},
  \newblock Chin.\ Phys.\ C {\bf 40}, 042001 (2016).
%
\bibitem{Krein:2016fqh} 
  G.~Krein,
  AIP Conf.\ Proc.\  {\bf 1701}, 020012 (2016).
%
\bibitem{Serna:2016kdb} 
  F.~E.~Serna, M.~A.~Brito and G.~Krein,
  \newblock AIP Conf.\ Proc.\  {\bf 1701}, 100018 (2016).
%
\bibitem{Nambu:1961tp} 
Y.~Nambu and G.~Jona-Lasinio,
\newblock Phys. Rev. {\bf 122}, 345 (1961).
%
\bibitem{Vogl:1991qt}
U.~Vogl and W.~Weise,
\newblock Prog. Part. Nucl. Phys. {\bf 27}, 195 (1991).
%
\bibitem{Klevansky:1992qe}
S.~Klevansky,
\newblock Rev. Mod. Phys. {\bf 64}, 649 (1992).
%
\bibitem{Hatsuda:1994pi}
T.~Hatsuda and T.~Kunihiro,
\newblock Phys. Rept. {\bf 247}, 221 (1994).
%
\bibitem{Bijnens:1995ww}
J.~Bijnens,
\newblock Phys. Rept. {\bf 265}, 369 (1996).
%
\bibitem{GutierrezGuerrero:2010md}
L.~X. Guti{\'e}rrez-Guerrero, A.~Bashir, I.~C. Clo\"et and C.~D. Roberts,
\newblock Phys. Rev. C {\bf 81}, 065202 (2010).
%
\bibitem{Krein:1990sf} 
  G.~Krein, C.~D.~Roberts and A.~G.~Williams,
  \newblock Int.\ J.\ Mod.\ Phys.\ A {\bf 7}, 5607 (1992).
%
\bibitem{Fischer:2003rp} 
  C.~S.~Fischer and R.~Alkofer,
  \newblock Phys.\ Rev.\ D {\bf 67}, 094020 (2003).
%
\bibitem{Ebert:1996vx} 
  D.~Ebert, T.~Feldmann and H.~Reinhardt,
\newblock  Phys.\ Lett.\ B {\bf 388}, 154 (1996).
%
\bibitem{Munczek:1994zz}
H.~J. Munczek,
\newblock Phys. Rev. D {\bf 52}, 4736 (1995).

\bibitem{Bender:1996bb}
A.~Bender, C.~D. Roberts and L.~von Smekal,
\newblock Phys. Lett. B {\bf 380}, 7 (1996).

\bibitem{Roberts:2010rn}
H.~L.~L. Roberts, C.~D. Roberts, A.~Bashir, L.~X. Guti{\'e}rrez-Guerrero and  P.~C. Tandy,
\newblock Phys. Rev. C {\bf 82}, {\mbox{065202}} (2010).

\bibitem{Roberts:2011wy}
H.~L.~L. Roberts, A.~Bashir, L.~X. Guti{\'e}rrez-Guerrero, C.~D. Roberts and
D.~J. Wilson,
\newblock Phys. Rev. C {\bf 83}, 065206 (2011).

\bibitem{Chen:2012qr}
C.~Chen, L.~Chang, C.~D. Roberts, S.-L. Wan and D.~J. Wilson,
\newblock Few Body Syst. {\bf 53}, 293 (2012).
%
\bibitem{Wilson:2011aa}
D.~J. Wilson, I.~C. Clo{\"e}t, L.~Chang and C.~D. Roberts,
\newblock Phys. Rev. C {\bf 85}, 025205 (2012).
%
\bibitem{Chen:2012txa}
C.~Chen, L.~Chang, C.~D. Roberts, S.~M. Schmidt, S.~Wan and D.~J. Wilson, 
\newblock Phys. Rev. C {\bf 87}, 045207 (2013).

\bibitem{Wang:2013wk}
K.-L. Wang, Y.-X. Liu, L.~Chang, C.~D. Roberts and S.~M. Schmidt,
\newblock Phys. Rev. D {\bf 87}, 074038 (2013).

\bibitem{Roberts:2011cf}
H.~L.~L. Roberts, L.~Chang, I.~C. Clo{\"e}t and C.~D. Roberts,
\newblock Few Body Syst. {\bf 51}, 1 (2011).

\bibitem{Segovia:2013rca} 
  J.~Segovia, C.~Chen, C.~D.~Roberts and S.~Wan,
  \newblock Phys.\ Rev.\ C {\bf 88}, 032201 (2013).

\bibitem{Segovia:2013uga} 
  J.~Segovia, C.~Chen, I.~C.~Clo{\"e}t, C.~D.~Roberts, S.~M.~Schmidt and S.~Wan,
  \newblock Few Body Syst.\  {\bf 55}, 1 (2014).
  
  \bibitem{Bedolla:2015mpa}
  Marco~A.~Bedolla, J.~J.~Cobos-Mart\'inez, and Adnan Bashir. Phys. Rev. D {\bf 92}, 054031 (2015).
  \newblock Phys.\ Rev.\ D {\bf 92}, 054031 (2015).
  
  \bibitem{Bedolla:2016yxq} 
  M.~A.~Bedolla, K.~Raya, J.~J.~Cobos-Mart\'inez and A.~Bashir,
  \newblock Phys.\ Rev.\ D {\bf 93}, 094025 (2016).
  
%
\bibitem{Battistel:2008fd}
  O.~A.~Battistel, G.~Dallabona and G.~Krein,
  \newblock Phys.\ Rev.\ D {\bf 77} (2008) 065025.
  
\bibitem{Battistel:2013cja} 
  O.~A.~Battistel and G.~Dallabona,
  \newblock Phys.\ Rev.\ D {\bf 80}, 085028 (2009); {\em ibid} 
  Phys.\ Rev.\ D {\bf 94}, 085011 (2016).

%
\bibitem{Battistel:2003gn}
  O.~A.~Battistel and G.~Krein,
  Mod.\ Phys.\ Lett.\ A {\bf 18} (2003) 2255.
%
\bibitem{Farias:2006cs} 
  R.~L.~S.~Farias, G.~Dallabona, G.~Krein and O.~A.~Battistel,
  \newblock Phys.\ Rev.\ C {\bf 77}, 065201 (2008).

\bibitem{Farias:2005cr} 
  R.~L.~S.~Farias, G.~Dallabona, G.~Krein and O.~A.~Battistel,
  \newblock Phys.\ Rev.\ C {\bf 73}, 018201 (2006).

\bibitem{Farias:2007zz} 
  R.~L.~S.~Farias, G.~Krein, G.~Dallabona and O.~A.~Battistel,
  \newblock Nucl.\ Phys.\ A {\bf 790}, 332 (2007).
%
\bibitem{Farias:2016let} 
  R.~L.~S.~Farias, D.~C.~Duarte, G.~Krein and R.~O.~Ramos,
  \newblock Phys.\ Rev.\ D {\bf 94}, 074011 (2016).
%
\bibitem{Batt-thesis}  O.~A Battistel, 
\newblock \textit{PhD Thesis 1999}, Universidade
Federal de Minas Gerais, Brazil (unpublished).
%
\bibitem{Sampaio:2002ii} 
  M.~Sampaio, A.~P.~Baeta Scarpelli, B.~Hiller, A.~Brizola, M.~C.~Nemes and S.~Gobira,
  Phys.\ Rev.\ D {\bf 65}, 125023 (2002).
%
\bibitem{Collins:1984xc} 
  J.~C.~Collins,
  {\em Renormalization : An Introduction to Re\-norm\-alization, The Renormalization Group, 
  and the Operator Product Expansion} (Cambridge University Press, Cambridge, 1984).
%
\bibitem{blaschke}
D.~Blaschke, P.~Costa, and Yu.~L.~Kalinovsky,
\newblock Phys.\ Rev.\ D {\bf 85}, 034005 (2012).
%
%
\bibitem{El-Bennich:2016qmb} 
  B.~El-Bennich, G.~Krein, E.~Rojas and F.~E.~Serna,
  \newblock Few Body Syst.\  {\bf 57}, 955 (2016).
%
\bibitem{Casalbuoni:2003cs} 
  R.~Casalbuoni, R.~Gatto, G.~Nardulli and M.~Ruggieri,
  \newblock Phys.\ Rev.\ D {\bf 68}, 034024 (2003)
  
%
\bibitem{ElBennich:2008xy} 
  B.~El-Bennich, O.~Leitner, J.-P.~Dedonder and B.~Loiseau,
  Phys.\ Rev.\ D {\bf 79}, 076004 (2009).
  
\bibitem{daSilva:2012gf} 
  E.~O.~da Silva, J.~P.~B.~C.~de Melo, B.~El-Bennich and V.~S.~Filho,
  Phys.\ Rev.\ C {\bf 86}, 038202 (2012).
 
 \bibitem{ElBennich:2012ij} 
  B.~El-Bennich, J.~P.~B.~C.~de Melo and T.~Frederico,
  Few Body Syst.\  {\bf 54}, 1851 (2013).
  
  \bibitem{deMelo:2014gea} 
  J.~P.~B.~C.~de Melo, K.~Tsushima, B.~El-Bennich, E.~Rojas and T.~Frederico,
  Phys.\ Rev.\ C {\bf 90}, 035201 (2014).
  
 \bibitem{Yabusaki:2015dca} 
  G.~H.~S.~Yabusaki, I.~Ahmed, M.~A.~Paracha, J.~P.~B.~C.~de Melo and B.~El-Bennich,
  Phys.\ Rev.\ D {\bf 92}, 034017 (2015).
 
\bibitem{ElBennich:2008qa} 
  B.~El-Bennich, J.~P.~B.~C.~de Melo, B.~Loiseau, J.-P.~Dedonder and T.~Frederico,
  Braz.\ J.\ Phys.\  {\bf 38}, 465 (2008).
 
 \bibitem{ElBennich:2009vx} 
  B.~El-Bennich, M.~A.~Ivanov and C.~D.~Roberts,
  \newblock Nucl.\ Phys.\ Proc.\ Suppl.\  {\bf 199}, 184 (2010).
%
\bibitem{Maris-Roberts}
P. Maris and C. D. Roberts, 
\newblock  Phys. Rev. {\bf C 56}, 3369
(1997).
%
\bibitem{PDG}
C. Patrignani et al. (Particle Data Group), 
\newblock Chin. Phys. C {\bf 40}, 100001 (2016). 
%
\bibitem{Maris:1999nt} 
  P.~Maris and P.~C.~Tandy,
  Phys.\ Rev.\ C {\bf 60}, 055214 (1999).
%
\bibitem{Becirevic:2012ti} 
  D.~Becirevic, V.~Lubicz, F.~Sanfilippo, S.~Simula and C.~Tarantino,
  JHEP {\bf 1202}, 042 (2012).
%
\bibitem{Haidenbauer:2007jq} J. Haidenbauer, G. Krein, U.-G. Mei{\ss}ner and A. Sibirtsev,
  Eur. Phys. J. A\, {\bf 33}, 107 (2007).
%
\bibitem{Haidenbauer:2008ff} 
  J.~Haidenbauer, G.~Krein, U.-G.~Mei{\ss}ner and A.~Sibirtsev,
  Eur.\ Phys.\ J.\ A {\bf 37}, 55 (2008) .
%
\bibitem{Haidenbauer:2010ch}
  J.~Haidenbauer, G.~Krein, U.-G.~Mei{\ss}ner and L.~Tolos,
  Eur.\ Phys.\ J.\ A {\bf 47}, 18 (2011).
%
\bibitem{Fontoura:2012mz} 
  C.~E.~Fontoura, G.~Krein and V.~E.~Vizcarra,
  Phys.\ Rev.\ C {\bf 87}, 025206 (2013).
%
\bibitem{Tsushima:1998ru} 
  K.~Tsushima, D.~H.~Lu, A.~W.~Thomas, K.~Saito and R.~H.~Landau,
  Phys.\ Rev.\ C {\bf 59}, 2824 (1999).
%
\bibitem{GarciaRecio:2010vt}
  C.~Garcia-Recio, J.~Nieves, and L.~Tolos,
  Phys.\ Lett.\ B {\bf 690}, 369 (2010).
%
\bibitem{GarciaRecio:2011xt}
  C.~Garcia-Recio, J.~Nieves, L.~L.~Salcedo, and L.~Tolos,
  Phys.\ Rev.\ C {\bf 85}, 025203 (2012).
%
\bibitem{Ko:2000jx} 
  S.~H.~Lee and C.~M.~Ko,
  Phys.\ Rev.\ C {\bf 67}, 038202 (2003).
%
\bibitem{Krein:2010vp} 
  G.~Krein, A.~W.~Thomas and K.~Tsushima,
  Phys.\ Lett.\ B {\bf 697}, 136 (2011).
%
\bibitem{Tsushima:2011kh} 
  K.~Tsushima, D.~H.~Lu, G.~Krein and A.~W.~Thomas,
  \newblock Phys.\ Rev.\ C {\bf 83}, 065208 (2011).
%
\bibitem{Krein:2012lra} 
  G.~Krein,
  PoS ConfinementX {\bf }, 144 (2012).
%
\bibitem{ElBennich:2011py} 
  B.~El-Bennich, G.~Krein, L.~Chang, C.~D.~Roberts and D.~J.~Wilson,
  Phys.\ Rev.\ D {\bf 85}, 031502 (2012)
%
\bibitem{Krein:2014vma} 
  G.~Krein,
  EPJ Web Conf.\  {\bf 73}, 05001 (2014).
%
\bibitem{El-Bennich:2016bno} 
  B.~El-Bennich, M.~A.~Paracha, C.~D.~Roberts and E.~Rojas,
  Phys.\ Rev.\ D {\bf 95}, no. 3, 034037 (2017).
%
\bibitem{Ballon-Bayona:2017bwk} 
  A.~Ballon-Bayona, G.~Krein and C.~Miller,
  arXiv:1702.08417 [hep-ph].
%
\bibitem{GhoshSernaKrein}
S. Ghosh, F. E. Serna, B. F. Inchausp, S. K. Das, G. Krein, in preparation.
%
\end{thebibliography}
\end{document}